\documentclass{vgtc}                          

\ifpdf
  \pdfoutput=1\relax                   
  \usepackage{graphicx}                
  \DeclareGraphicsExtensions{.pdf,.png,.jpg,.jpeg} 
\else
  \usepackage{graphicx}                
  \DeclareGraphicsExtensions{.eps}     
\fi%

\graphicspath{{figures/}{pictures/}{images/}{./}} 

\usepackage{microtype}                 
\PassOptionsToPackage{warn}{textcomp}  
\usepackage{textcomp}                  
\usepackage{mathptmx}                  
\usepackage{times}                     
\usepackage{cite}                      
\usepackage{tabu}                      
\usepackage{booktabs}                  
\usepackage[table]{xcolor}
\usepackage{enumitem}

\definecolor{darkblue}{RGB}{107,174,214}
\definecolor{lightblue}{RGB}{198,219,239}

\definecolor{darkgreen}{RGB}{116,196,118}
\definecolor{lightgreen}{RGB}{199,233,192}

\definecolor{darkorange}{RGB}{253,141,60}
\definecolor{lightorange}{RGB}{253,208,162}

\definecolor{darkpurple}{RGB}{158,154,200}
\definecolor{lightpurple}{RGB}{188,189,220}

\definecolor{darkred}{RGB}{251,99,79}
\definecolor{lightred}{RGB}{254,198,190}

\definecolor{darkpink}{RGB}{248,112 ,151}
\definecolor{lightpink}{RGB}{252,196,212}

\onlineid{0}

\vgtccategory{Research}

\vgtcinsertpkg

\title{Characterizing Uncertainty in the Visual Text Analysis Pipeline}

\author{ 
Pantea Haghighatkhah $^\diamondsuit$ 
\and Mennatallah El-Assady $^\clubsuit$ 
\and Jean-Daniel Fekete $^\spadesuit$
\and Narges Mahyar $^\ddag$ %
\and Carita Paradis $^\S$
\and Vasiliki Simaki $^\S$%
\and Bettina Speckmann$^\diamondsuit$\\ %
     } %

\affiliation{$\diamondsuit$ TU Eindhoven $\clubsuit$ AI Center, ETH Zürich \\$\spadesuit$ Inria \& Université Paris-Saclay $\ddag$ University of Massachusetts Lowell $\S$ Lund University}
\teaser{
 \vspace{-1em}\includegraphics[width=\textwidth]{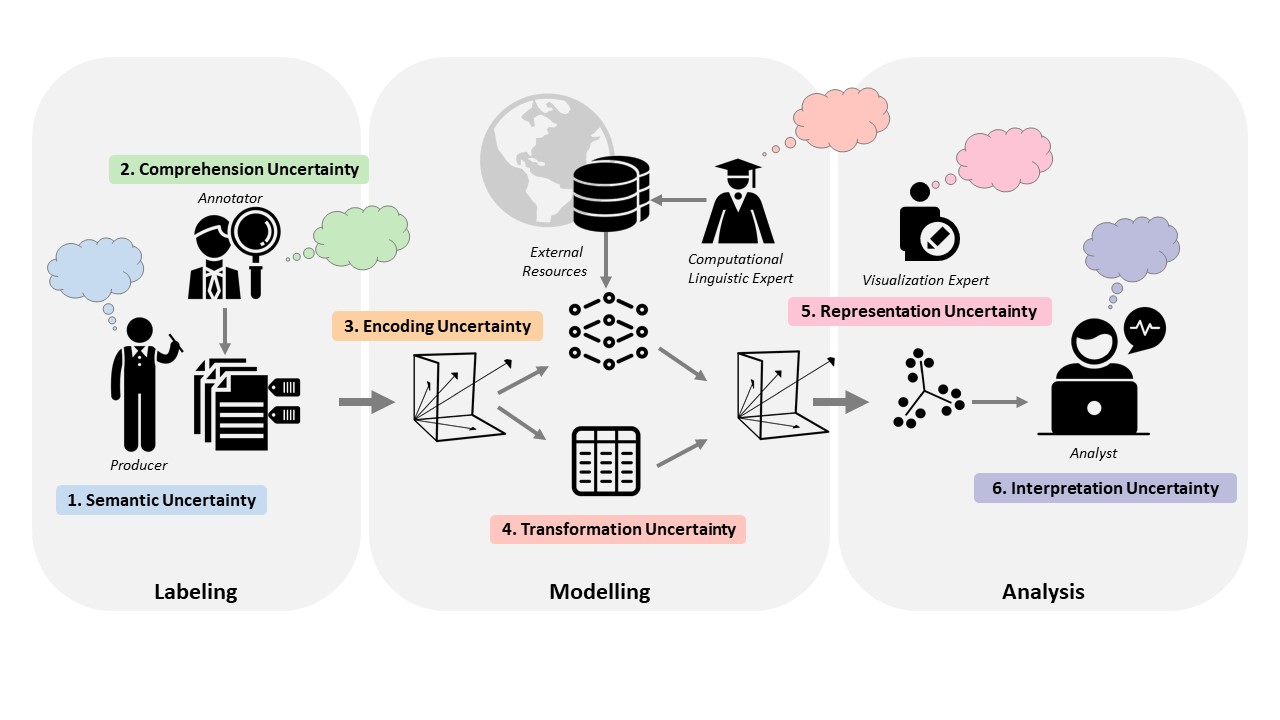}\vspace{-4em}
 \caption{We characterize six sources of uncertainty in the three stages of the visual text analysis  pipeline. %
 }\label{fig:teaser}
}

\abstract{Current visual text analysis approaches rely on sophisticated processing pipelines. Each step of such a pipeline potentially amplifies any uncertainties from the previous step. 
To ensure the comprehensibility and interoperability of the results, it is of paramount importance to clearly communicate the uncertainty not only of the output but also within the pipeline. 
In this paper, we characterize the sources of uncertainty along the visual text analysis pipeline. Within its three phases of labeling, modeling, and analysis, we identify six sources, discuss the type of uncertainty they create, and how they propagate. 
The goal of this paper is to bring the attention of the visualization community to additional types and sources of uncertainty in visual text analysis and to call for careful consideration, highlighting opportunities for future research. 
} 

\CCScatlist{
  \CCScatTwelve{Human-centered computing}{Visu\-al\-iza\-tion}{Visu\-al\-iza\-tion techniques}{Treemaps};
  \CCScatTwelve{Human-centered computing}{Visu\-al\-iza\-tion}{Visualization design and evaluation methods}{}
}

\begin{document}


\firstsection{Introduction}

\maketitle

With the rapid evolution of natural language processing (NLP) supported by machine learning and other technologies, it becomes possible to analyze and explore large and diverse textual corpora. Visualization is frequently used to support these analyses and explorations. 
We call the general domain connecting text processing, analysis, and visualization \emph{Visual Text Analysis}~\cite{liu2018bridging}.
Visual text analysis is used by linguistic experts, domain experts in the social sciences and humanities, or the general public. It is used for various tasks, such as providing an overview, searching, finding related documents, or understanding stylistic patterns within documents, to name a few. All the analytical and exploratory tasks are done after the text has been processed through multiple steps, as shown in \autoref{fig:teaser}. 
Each of the steps may, in turn, introduce some level of uncertainty. However,  few text visualization systems take into account these multiple uncertainties, in particular,  few text visualizations provide cues about them.\\

Uncertainty may come from different sources, it refers to epistemic situations involving imperfect or unknown information. 
Uncertainty and error are often used interchangeably. 
For the purpose of this paper, we consider errors to be deliberate and systemic; they require manual intervention in the processing pipeline to be corrected~\cite{beck2020representation}.
If uncertainty is caused by errors in the data, then we generally have no control over it.
If the uncertainty is caused by our processing pipeline, then we might or might not be able to precisely identify its source and contribution. 
In the visual text analysis pipeline, certain processing steps, such as dimensionality reduction, may cause artifacts.
These artifacts usually imply that the (visual) representation is not faithful to the underlying data anymore, and as such, they introduce a type of uncertainty. 
As opposed to errors, artifacts are the result of a deliberate decision by the analyst and can be (automatically) undone by choosing another processing method. Frequently, though, all methods at the analyst's disposal will introduce some sort of artifacts, which can only be mitigated via careful trade-offs.\\

The main contribution of our paper is a characterization of the multiple sources of uncertainty in the visual text analysis pipeline (see \autoref{sec:uncertainty}).
Not reporting uncertainty generates confusion and causes lack of trust for the users~\cite{Trust}. However, traditional methods for visualizing uncertainty (e.g. \cite{UncertaintyVis}) might not always apply to the variety of issues that appear and propagate through the text analysis pipeline. 
We hope that our paper will contribute to a better understanding of the uncertainty in the analysis pipeline and motivate future research. 


\section{Related Work}



In this section, we briefly discuss prior work on uncertainty modeling and research on uncertainty visualization. We then describe the 
importance of modeling and communicating uncertainty in text data. 

\subsection{Uncertainty Modeling}

Uncertainty, error, ambiguity, and other representation issues have been modeled in various ways across research communities. Specifically for linguistic processing and visual analysis, it is imperative to capture such issues for every task at hand. For example, for the task of linguistic annotation, modeling ambiguities and inter-annotator agreement is of utmost importance, as described by Beck et al.~\cite{beck2020representation}. For other tasks, such as sentiment analysis,  Bayesian deep learning can be utilized to characterize and model uncertainty~\cite{xiao2019quantifying}. For the task of multi-labeled text classification,  Chen et al.~\cite{chen2020uncertainty} quantify uncertainty in the transformation step of the pipeline, deploying experiments to measure both \textit{aleatoric} and \textit{epistemic} uncertainty (various natures of uncertainty). Other approaches focus on uncertainty communication, for example through visualization. For example, Collins et al.~\cite{collins2007visualization} model the uncertainty of text data during visualization using lattice graphs. It's a specific graph-based visualization that expresses the multiple possible outputs that are hidden from the users.
While there are many uncertainty modeling approaches, in the context of this paper, we focus on the overall visual text analysis pipeline, including uncertainty visualization, as described in the next subsection.

\subsection{Uncertainty Visualization}
 According to Zuk and Carpendale, uncertainty is a fundamental part of any analytic or reasoning process \cite{zuk2007visualization}. 
Identifying and communicating uncertainty is critical for many analytical tasks. Prior work has addressed the importance of identifying, quantifying, and visualizing uncertainty (e.g., \cite{ hullman2019authors, kay2016ish}). The complexity of visualizing uncertainty is a known challenge \cite{greis2017designing}. Researchers in visualization have investigated various techniques for conveying uncertainty through interaction, animation, and sonification \cite{tse2016we}. Some examples include Value Suppressing Uncertainty Palettes (VSUPs), which adjust the visual channel allocated to uncertainty based on the level of uncertainty
\cite{correll2018value}, and Hypothetical Outcome Plots (HOPs), that animates a finite set of individual draws \cite{hullman2015hypothetical}.

\subsection{The Importance of Uncertainty in Text Data}
While previous work explores various methods for visualizing uncertainty in general, due to the complexity and ambiguity involved in text data, traditional methods might not always apply to the variety of issues that appear and propagate through the text analysis pipeline. Recent work in text analytics suggests that text visualization needs more careful consideration of uncertainty, its sources, and potential ways to visualize them \cite{hofman2020visualizing}. 

The uncertainty in text visualization comes from many origins. First, we need to consider that text is an imperfect representation of human thoughts; therefore, encoding thoughts in a text by nature produces artifacts. Another issue is that people might have a different understanding and interpretation of any given text. Hence, one single text input can result in multiple interpretations, which affect not only the interpretation of the receiver but also the annotators, because they might have various interpretations. 

Uncertainty can hamper judgment in any decision-making scenario \cite{tversky1974judgment}, but the impact of uncertainty exacerbates in domains where text data is utilized for important decisions, such as civic decision-making~\cite{Baumer2022OfCourse}, or in humanities research where historians and archivists analyze the text data in newspaper archives to answer
fundamental questions about society~\cite{handler2022clioquery}.

The open challenge in the text visual analytics domain is how to identify sources of errors and artifacts in the text visualization pipeline and how to design techniques and embed them in various stages of the pipeline to communicate uncertainty to various actors such as annotators and end users. 
Due to the inherent complexity of uncertainty visualization, another open challenge is how to quantify, model, and visualize the uncertainty to ensure users can understand and interpret it correctly.

\section{Uncertainties in Visual Text Analytics Pipelines}\label{sec:uncertainty}

Uncertainty in text visualization stems from many origins.  
\autoref{fig:teaser} visually summarizes the six sources of uncertainty we identified in the three stages of the visual text analysis pipeline: labeling, modeling, and analysis.
In the following, we describe the different types of uncertainty in detail. 

\subsection{Semantic Uncertainty} 
At the start of the pipeline, uncertainty can be introduced by the text \emph{producer}, that is, by the way in which a person types a text or adds a text document to a collection. The errors can be related to the text itself or to the metadata, such as the date of the text production or the text attribution (author). Transcription from a manuscript or from an oral source can always contain transcription errors or misspelled names that will propagate errors and hence uncertainty down the pipeline. Note that we discuss here a type of uncertainty that comes from the producer, and was not caused by textual (e.g misspellings) and metadata errors.

\begin{table}[h]
    \renewcommand{\arraystretch}{1.3}
        \begin{tabular}{p{8cm} }
        \cellcolor{darkblue}\textbf{Semantic Uncertainty }\\
         \cellcolor{lightblue} The uncertainty caused by the producer's mental linguistic model and translation of thought to text. This uncertainty is also caused by the expression of the speaker's feeling that something is not known or certain, often conveyed by means of specific linguistic markers in the text.
    \end{tabular}
\end{table}

\noindent
Text is an imperfect representation of human thoughts, therefore encoding thoughts in a text by nature produces artifacts. \autoref{fig:teaser} shows that the first type of uncertainty concerns the origin of what is communicated through language. This uncertainty refers to the fact that expressions of language are recontextualized thoughts, events, and experiences through language in the written medium. In the transition from the producer through language to the reader, information may or may not be lost. We refer to this as \textit{semantic uncertainty} on the part of the producer. This type of uncertainty cannot always be detected in the text as the producer uses their own linguistic model that can differ from others. There are cases where the semantic uncertainty can be observed from the text with the speaker using language to express ``\textit{doubt as to the likelihood or truth of what she or he is saying}''~\cite{simaki2020annotating}, their opinions, facts, or ideas. If producers pay attention to this problem, they often indicate this through expressions such as \textit{may}, \textit{not sure}, \textit{might}, \textit{could}. \\

\noindent \textbf{Examples}: \textit{``I am not sure how to get there.'', ``We might go to the restaurant.'', ``We have enough time, haven’t we?.''}
In these examples, the producers use specific linguistic items (\textit{not sure}, \textit{might}) and ways (tag question \emph{haven't we?}) to express their doubt/uncertainty on the particular way to get to a location, the possibility to go to a restaurant, or whether there is enough time to do something.  If no such expressions of uncertainty are used, then one may not be able to identify the semantic uncertainty of the text.

\noindent \textbf{Challenges}: The main challenge is to produce accurate and precise representations in the pipeline of the refined semantic relations expressed in the texts, as well as to identify the uncertainty in the content and how this affects the overall meaning of the text.

\subsection{Comprehension Uncertainty} 
The second source of uncertainty comes from data capturing and annotation performed on the text. This step can include entity recognition and more general annotations such as sentiment analysis, introduced by \textit{annotators}. This enrichment adds higher-level semantics, but also possible errors or misinterpretations that might change the result of the annotation and introduce uncertainty.

\begin{table}[h]
    
    \renewcommand{\arraystretch}{1.2}
        \begin{tabular}{p{8cm} }
        \cellcolor{darkgreen}\textbf{Comprehension Uncertainty}\\
         \cellcolor{lightgreen} The uncertainty caused during the data collection and annotation process due to technical challenges, perceptual differences and/or other limitations.  
    \end{tabular}
\end{table}

\noindent
People do not respond in the same way to a given text. Hence, one single text input can result in multiple interpretations. This brings us to the second type of uncertainty that we identified, namely, \emph{comprehension uncertainty}. This uncertainty can be caused during (i) the data capturing and collection process, and (ii) the data annotation process. In (i), the uncertainty is caused by issues related to the representativeness of various text genres/types/contents/topics and their balance in the data set, the noise that irrelevant content creates in the data set, and various biases in the data extraction and collection process. In (ii), multiple factors contribute to the uncertainty, as human annotators come to the task with different linguistic understandings and perceptions of the text. One single text input can result in multiple interpretations, and hence, different decisions by annotators. Phenomena such as polysemy and ambiguity, and vague or generic annotation guidelines can further influence the uncertainty in the annotation process. Also, the annotators’ different perceptual systems play an important role in their final decision. All these factors can lead to a high level of disagreement between the annotators, which creates uncertainty about the reliability of the annotated data, and adds more uncertainty in the NLP pipeline.\\ 

\noindent \textbf{Examples}: Consider a case where the annotator is asked to identify the sense that is addressed or discussed in each sentence e.g. taste, touch, smell, etc. The annotator is given the sentence \textit{``The food is too soft for me.''}. It depends on the annotator's comprehension of the sentence to choose \textit{smell}, \textit{touch} (texture) or the \textit{taste} as the associated sense with this sentence. This case concerns the comprehension uncertainty caused by (ii).\\


\noindent \textbf{Challenges}: Apart from the various difficulties that may be faced during the data capturing part, a basic issue that needs to be considered is the design of a solid and well-crafted data collection and annotation protocol, as well as a thorough control (possibly automatic) before finalizing the annotated data set.
If this is not adequately addressed, less reliable data is produced, and as a result, more uncertainty is added to the pipeline.\\

\subsection{Encoding Uncertainty}
Text encoding in itself can also cause uncertainty. Some complex formats  have been designed to encode text in a rich yet faithful fashion, e.g., through the Text Encoding Initiative (TEI)~\cite{ide1995text}. However, very few projects use these sophisticated mechanisms. More often than not, visual text analysis projects resort to simpler encodings that can lead to  information loss or generate ambiguity that causes uncertainty. Even the TEI guidelines, although very rich and well documented, cannot avoid unexpected variations in their interpretation.

\begin{table}[h!]
    
    \renewcommand{\arraystretch}{1.2}
        \begin{tabular}{p{8cm} }
        \cellcolor{darkorange}\textbf{Encoding Uncertainty}\\
         \cellcolor{lightorange} The uncertainty caused by the data mapping to a data structure, which could lead to a lossy representation of the input.
    \end{tabular}
\end{table}
\noindent
The TEI guidelines~\cite{sperberg1994guidelines}, designed to define best practices to encode textual sources with a rich vocabulary of annotations, mention several mechanisms for encoding uncertainty, such as ``\textit{levels of certainty}'' and “precision” in the chapter ``Certainty, Precision, and Responsibility,'' and encoding for text segments such as ``\textit{unclear}'', ``\textit{gap}'' for the transcription of text or speech. All of these textual or linguistic uncertainties are idiosyncratic and intrinsic to our languages, texts, and speech structures. The TEI also allows encoding alternative interpretations for the same text segment (using the $<choice>$ element), as well as marking visible errors ($<sic>$) and possible corrections ($<corr>$). These annotations can become very rich and are currently not supported consistently by visualization systems; they are mostly ignored.\\

\noindent
The use of external resources to enrich the text is also a source of uncertainty. For example, when named entities are recognized in a text (e.g., a name or a place name), many NLP systems try to \emph{resolve} them, keeping a dictionary of mentioned persons, looking them up in popular databases (e.g., Wikipedia), or trying to find the location of a named place. These enrichments also lead to errors and consequently to uncertainty. For example, the exact address of a person in the 18th century might not be resolved accurately by a modern geocoding service when the street name has changed; the address is then resolved as a city instead of a precise block location.\\

\noindent \textbf{Example}: \textit{One-hot-vector} encoding of the sentence: 
``\textit{This is an example sentence that contains the word example.}''\\ \vspace{-1em}
\begin{itemize}[nosep]
    \item this: $<1,0,0,0,0,0,0,0,0>  $
    \item is: $<0,1,0,0,0,0,0,0,0>  $
    \item an: $<0,0,1,0,0,0,0,0,0>  $
    \item example: $<0,0,0,2,0,0,0,0,0>  $
    \item ...
\end{itemize}
\noindent 
Using a \textit{one-hot-vector} encoding as illustrated above is an efficient way to gather statistical information about the text, but leads to the loss of the word contexts, as the order is not preserved. With such an encoding, uncertainty cannot be reverted, as the original sentence cannot be reconstructed by reverting the encoding. \\

\noindent \textbf{Challenges}:  The main challenge for encoding is to avoid loss of information if this information is usable in the processing. However, some loss is unavoidable since the text data needs to be structured in a form that can be processed. Measuring the information loss during this step can enable efficient communication of encoding uncertainty. This builds an interesting area for future research.

\subsection{Transformation Uncertainty} 
The encoding typically represents the data as embedding vectors in a high-dimensional space. These get transformed through NLP models that are either exclusively considering the internal data from the pipeline or additionally rely on external resources, such as language modeling or externalizing expert knowledge and feedback. The NLP models introduce transformation uncertainty into the pipeline. Such transformations often rely on design decisions by the \textit{computational linguistics experts}.

\begin{table}[h]
    
    \renewcommand{\arraystretch}{1.2}
        \begin{tabular}{p{8cm} }
        \cellcolor{darkred}\textbf{Transformation Uncertainty}\\
         \cellcolor{lightred} Uncertainty that is introduced through computations, for example, through language modeling or injection of expert knowledge and feedback.
    \end{tabular}
\end{table}

\noindent 
After encoding and possible enrichment, the text is often transformed to be easy to analyze. Most search engine will transform sentences or documents into high-dimensional vectors using language modeling approaches, such as \textit{word2vec}, \textit{doc2vec}, \textit{BERT}, or \textit{GPT-3}. These vectors allow finding similar documents fast, but they also abstract-out the text and turn it into a representation that humans cannot interpret directly. These transformations are complex and can generate artifacts, errors, and lead to uncertainty.\\

\noindent \textbf{Example}: ``\textit{This sentence is about \textbf{cats}}.'' and ``\textit{This sentence is not about \textbf{cats}, but about dogs}.''  If we consider the two sentences above and a model that is mapping them to different topics~\cite{el2019lingvis}, transformation uncertainty can arise if the word \textit{cats} in the second sentence causes it to be partly considered as belonging to the topic cats, ignoring the negation. This example is trivial, but more nuanced issues arise in many NLP pipelines.\\

\noindent \textbf{Challenges}:  Choosing the appropriate algorithm to model and transform text is crucial for visual text analysis. However, most linguistic models are not fine-tuned to the exact problems they are applied to. Hence, a major challenge is to configure the appropriate processing steps for a given text input and to allow for the capturing of uncertainties in each processing step.

\subsection{Representation Uncertainty} 
The output of the transformation step needs to be presented to a human by means of visualization. This is a potential source of uncertainty. For example, 2D visualization of high-dimensional vectors that are trained to represent documents is prone to dimensionality reduction artifacts, namely, distortions. This can cause uncertainty about actual distances between vectors (documents) in the original space. These uncertainties are usually introduced by the \textit{visualization experts}.

\begin{table}[h]
    
    \renewcommand{\arraystretch}{1.2}
        \begin{tabular}{p{8cm} }
        \cellcolor{darkpink}\textbf{Representation Uncertainty}\\
         \cellcolor{lightpink} In order to visualize the high-dimensional vectors resulting from the transformation step of the pipeline, one needs to use projection techniques to create 2D/3D representation of these vectors. Such a process generates artifacts and errors in the distances that cause Representation Uncertainty.
    \end{tabular}
\end{table}
\noindent 
The output of the transformation step is often a high-dimensional vector that embeds certain information about words, sentences, or documents within itself. Let us assume that in our example pipeline, the transformation step outputs vectors that represent documents. The distances between the document vectors capture the relatedness of the corresponding documents. Therefore, visual observation of the document vectors can be a valuable tool for the purpose of the topic, genre, or in general document analysis. In order to visualize such vectors, one has to reduce the dimensionality of the vectors to two or three dimensions using projection techniques such as t-SNE \cite{tsne} or UMAP \cite{umap}. Projecting high-dimensional data to lower dimensions is inherently lossy and can cause distortions. Hence, the resulting visualization of the low-dimensional projection may be unfaithful to the original distances between the document vectors. Therefore, the 2D/3D representation may carry errors and artifacts that result in \textit{representation uncertainty}.\\

In general, when two vectors are close, their corresponding documents are thought to be similar. However, the low-dimensional representation may include seemingly coherent clusters that do not exist in the original data (false neighbors) or conversely miss existing clusters of the original vectors due to topological artifacts of the projection (missing neighbors). \\

A visualization should at least inform the users of these possible artifacts, preferably indicate areas they occur and, if possible, allow resolving them. The lack of communication regarding representation uncertainty may hinder the trust of the user. Currently, few visualization systems inform users of possible artifacts, and almost none provide techniques to overcome them, especially for many data points. Addressing that problem is essential for text visualization in particular, but also for multidimensional visualization in general. Aupetit~\cite{topo_aupetit,interactive_JDF} has proposed a few techniques for small amounts of data and Martins \cite{MARTINS201426} for larger amounts, but they are targeted towards expert users.\\

\noindent \textbf{Example}: The artifacts of false neighbors and missing neighbors in the 2D representation can be considered as examples of contributing factors to the representation uncertainty.\\

\noindent \textbf{Challenges}:  The most significant challenge is to resolve errors and artifacts generated by this part of the pipeline. However, this is not always achievable due to the lossy nature of the dimensionality reduction methods. Therefore, another interesting task is quantifying errors and artifacts and visualizing them in the representation step. The key here is to create an intuitive yet expressive visualization for communication to reduce the representation uncertainty.

\subsection{Interpretation Uncertainty} At the last stage of the pipeline the receiver or \textit{analyst} inspecting the visualization is interpreting the data through the lens of the visual design; interpretation uncertainty is due to the mindset of the \textit{analyst} (user).

\begin{table}[ht]
    
    \renewcommand{\arraystretch}{1.2}
        \begin{tabular}{p{8cm} }
        \cellcolor{darkpurple}\textbf{Interpretation Uncertainty}\\
         \cellcolor{lightpurple} The uncertainty caused by the analyst's interpretation of the text visualization at the last stage of the pipeline is called Interpretation Uncertainty. \\
    \end{tabular}
\end{table}

\noindent
The representation stage of the pipeline provides the analyst with a visualization of the input text. The analyst has their own linguistic model that they use to interpret the text data. This model is unique to each user (analyst). Besides this, the user has a personal interpretation of the observed visualization that can differ from other humans' interpretations. This can be due to different interpretations of long and short distances and also what counts as scattered and what counts as a coherent cluster. Similar uncertainties can occur when the analyst is interpreting the colors in the visualization. 
Both the linguistic model and the visual interpretation of the analyst contribute to the uncertainty generated at the last stage of the pipeline.\\

\noindent \textbf{Examples} Analyst may interpret a set of points associated with text as a \textit{coherent cluster} or conversely interpret it as a \textit{scattered point set}.\\

\noindent \textbf{Challenges}:  
Text data in nature is ambiguous, and therefore there is no single canonical data representation or universally accepted interpretation \cite{Poesio2005Reliability,Pavlick2019Inherent}. As a result, the human who reads the visualization and multiple representations of the text can  misinterpret the representations.
A challenge for the user of the text visualization is to come up with a suitable qualitative or quantitative metric supporting the interpretation of the visualization and therefore reach a more reliable interpretation of the visualization. 

Furthermore, the questions that the analyst seeks to answer using the text visualization should be designed such that they can be answered mostly by quantitative measures and numbers to minimize the effects of personal interpretations.

\section{Discussion \& Conclusion}

In this paper, we presented a detailed description of the uncertainty surrounding each step of the visual text analysis pipeline. 
We did, however, restrict ourselves to the uncertainty of each step in isolation, and the impact of each of the identified uncertainty on future processing steps remains unclear. The steps of the text visualization pipeline can be more intertwined in reality, for instance, the visualization expert and the analyst may revisit the data and output of previous steps of the pipeline.
Hence, one of the main challenges, and an opportunity for future work, is to characterize the interaction between different steps of the pipeline and the propagation of uncertainty.
One plausible assumption is that the propagation  \textit{amplifies}  previous uncertainties, however, it can also \textit{nullify}  errors and artifacts, hiding possible issues that could lead to harmful rationalizations when interpreting the final results~\cite{SeEl2022Beware}. \\

As previous work suggests \cite{chen2020uncertainty}, the nature of uncertainty can be characterized as \textit{aleatoric} or \textit{epistemic}. Uncertainty in each step of the pipeline can have either of these natures. The first challenge is to identify the nature of uncertainty in each step.  
If we can identify the sources of uncertainty and their effects throughout the pipeline, the next major challenge is to (visually) communicate their impact.
Specifically, exposing all inner workings of the analysis steps and visualizing the errors per step directly, most likely leads to visual and cognitive overload.
Possible solutions might be to design in-situ and on-demand explanations through  \textit{co-adaptive analytics}~\cite{sperrle2021co} by learning from user interactions during analysis and teaching them potential impacts of uncertainty~\cite{sperrle2020learning}.

\acknowledgments{
We would like to thank the organizers of Dagstuhl Seminar 22191 \emph{``Visual Text Analytics''}\footnote{\url{https://www.dagstuhl.de/en/program/calendar/semhp/?semnr=22191}}. This work has benefited substantially from the discussions held during this seminar.}

\bibliographystyle{abbrv-doi}

\bibliography{template}
\end{document}